\documentclass[12pt,a4paper]{article}
\usepackage[margin=1in]{geometry}
\usepackage{graphics,epsfig}
\usepackage{color}
\usepackage{float}
\usepackage{tikz} 
\usepackage{booktabs}
\usepackage{multirow}
\usepackage[flushleft]{threeparttable}
\usepackage{fancybox}
\usepackage{lipsum}
\usepackage{eurosym}
\usepackage{euscript}
\usepackage{amsthm}
\usepackage{amsmath}
\usepackage{amsfonts}
\usepackage{amssymb}
\graphicspath{{./fig/}}
\usepackage{hyperref}
\usepackage{pdflscape}
\usepackage{charter}

\usepackage{subcaption}
\usepackage{tikz}
\usepackage{adjustbox}
\usetikzlibrary{arrows,positioning}
\usetikzlibrary{mindmap,trees,shadows,backgrounds}
\usepackage[all]{genealogytree}

\definecolor{deepblue}{RGB}{0,129,188}
\hypersetup{
bookmarks=true,
unicode=false,
pdftoolbar=true,
pdfmenubar=true,
pdffitwindow=false,
pdfstartview={FitH},
pdfnewwindow=true,
colorlinks=true,
linkcolor=deepblue,
citecolor=deepblue,
filecolor=black,
urlcolor=black,
pdfkeywords={Genealogy}}
\makeatother
\let\orgautoref\autoref

\renewcommand{\autoref}[1]
{\def\appendixname{}\def\equationautorefname{Equation}\def\figureautorefname{Figure}\def\subfigureautorefname{Figure}\def\tableautorefname{Table}\def\sectionautorefname{Section}\orgautoref{#1}}

\AtBeginDocument{\hypersetup{citecolor=deepblue, linkcolor=deepblue,filecolor=black,urlcolor=deepblue}}

\title{Internal Migrations in France in the Nineteenth Century\footnote{Research conducted within the `ACTINFO PROJECT' under the aegis of the Risk Foundation, a joint initiative by the GENES, the University of Rennes 1, the University of Paris-Est La Vall\'ee and COVEA. A preliminary work was presented during the `Science XXL' days organized in March 2017 at the French Institute for Demographic Studies (INED). We thank Olivier Cabrignac and Jérôme Galichon for their help on data exploration, as well as the participants for the discussions we had then, which motivated some of the elements presented in this study. We also thank the members of INED's History and Populations Unit for their comments. We benefited from fruitful discussions with the participants of the `UseR' conferences that were held in May 2018, the `R meeting' that took place in Rennes in July 2018, and the `XXIX International Biometric Conference' that were organized in Barcelona in July 2018.}}
\usepackage{authblk}
\author[a, b]{Arthur Charpentier}
\author[a,c]{Ewen Gallic\thanks{Corresponding author: \texttt{ewen.gallic@gmail.com}}}
\affil[a]{CREM, UMR CNRS 6211, 7 Place Hoche, 35065 Rennes Cedex, France.}
\affil[b]{Universit\'e de Rennes 1, 2 rue du Thabor -- CS 46510, 35042 Rennes Cedex, France.}
\affil[c]{Institut Louis Bachelier, Palais Brongniart 28 place de la Bourse 75002 Paris, France.}

\usepackage{fancyhdr}
\usepackage[utf8]{inputenc}
\usepackage{lmodern}
\usepackage[T1]{fontenc}
\usepackage[scaled]{helvet}
\usepackage[round]{natbib}
\usepackage[english]{babel}

\begin{document}

\maketitle

\begin{abstract}
	The digital age allows data collection to be done on a large scale and at low cost. This is the case of genealogy trees, which flourish on numerous digital platforms thanks to the collaboration of a mass of individuals wishing to trace their origins and share them with other users. The family trees constituted in this way contain information on the links between individuals and their ancestors, which can be used in historical demography, and more particularly to study migration phenomena. This article proposes to use the family trees of $238,009$ users of the Geneanet website, or $2.5$ million (unique) individuals, to study internal migration. The case of 19th century France is taken as an example. Using the geographical coordinates of the birthplaces of individuals born in France between 1800 and 1804 and those of their descendants, we study migration between generations at several geographical scales.  We start with a broad scale, that of the departments, to reach a much finer one, that of the cities. Our results are consistent with those of the literature traditionally based on parish or civil status registers. The results show that the use of collaborative genealogy data not only makes it possible to recover known facts in the literature, but also to enrich them.
\end{abstract}


\clearpage

%\tableofcontents

% ---------------------------------------- %
\section{Introduction}\label{intro}
% ---------------------------------------- %

Historical demography is a discipline founded in the second half of the 20th century aimed at giving an overview of the characteristics of the population in the past. The pioneering work of \citet{henry_1956_Population} has given rise to a multitude of studies on topics such as mortality \citep{blayo_1967_donnees,Blayo_1975_Population_mortalite,henry_1975_Population} and family structure \citep{matthijs_2010_HF}. Migration has also been widely studied, due to the increase in movements that have been observed over the past centuries. Several factors have been put forward to explain this growth in migration. They can be divided into two categories: the first involves the role of the family, and the second explains migration through social background and education.

Regarding the role of the family in migration decisions, \cite{kesztenbaum_2008_HF} showed the positive influence of siblings: having a brother who had already migrated has a positive impact on the migration of other brothers. However, the author notes that network effects play only a weak role in destination choices. This poor network effect is also reported by \cite{bonneuil_2008_SH}. Family composition also plays an important role. \cite{dribe_2003_HF} showed that the number of elderly individuals has a negative influence on the propensity to migrate, while conversely, the number of young individuals plays a positive role in the propensity to migrate.

Social environment and education are two other explanatory factors of migration highlighted in the literature. The research of \citet{heffernan_1989_literacy} has exposed the existence of a positive correlation between literacy and migration. This link was examined by \cite{bourdieu_2000_Annales}. These authors have shown that the higher the level of education, the more likely individuals are to migrate. \cite{rosental_2004_ADH} and \citet{bonneuil_2008_SH} obtained similar results. \cite{rosental_2004_ADH} also pointed to a distinction between urban and rural areas in the 19th century, with urban populations tending to be more attracted to cities and more willing to travel long distances.

Many empirical studies in historical demography, whether or not specifically concerned with migration phenomena, rely on register data. The information provided by the registers relates to the dates of births, marriages and deaths of individuals. In France, parish registers and civil registers since the French Revolution have been widely used, but are not the only source of information in the literature. For example, military registration numbers established by the military administration constitute a different source, used in particular to study migration during life \citep{ho_1971_Population,kesztenbaum_2008_HF,kesztenbaum2014etude}. A source closely linked to the existence of the various registers, since it relies on them, has contributed to the improvement of knowledge of the history of our ancestors: genealogical data. It is noteworthy that Mormons have built up extensive genealogical databases at the University of Utah that have contributed to population studies in the past \citep{bean_1978_HM,lindahl_2013_AE}. Recently, with the era of the development of computer techniques, this type of data seems to benefit from an increasing interest. The digital revolution makes it possible to access genealogical information in a relatively short period of time, at a lower cost and with less effort. Many websites offer their users to reconstitute their family trees. This is the case of \href{https://www.wikitree.com/}{wikitree.com}, \href{https://www.familysearch.org/}{familysearch.org} or \href{http://geni.com/}{geni.com}. So far, the information recorded by the users has mainly been used to study the longevity of individuals \citep{gavrilova_2007_NAAJ, gavrilov_2002_BSD,cummins_2017_JEH,fire_2015}.

Recently, \citet{Kaplanis_2018_Science} explored the family trees of several million individuals. Their study shows that genealogical data obtained through the collaboration of amateurs can produce high quality genealogical trees. We use the same type of data to study migrations, with an application to the French case in the 19th century. More specifically, we exploit information provided by amateur and professional genealogists who have built their family tree on a website called \href{https://en.geneanet.org/}{Geneanet}. Users of the site have given information about the places and dates of birth, marriage and death of their ancestors. Based on this information, we study the migration of descendants of individuals born at the beginning of the 19th century in France. The concept of migration can be understood in different ways, as recalled by \citet{Greenwood_1997}. One of them, proposed by the United Nations, states: `\textit{A migration is defined as a move from one migration defining area to another (or a move of some specified minimum distance) that was made during a given migration interval and that involved a change of residence}' \citep{UN_1970_Migration}. This definition involves two notions: a temporal one, and a spatial one. This distinction can be found in the classification into four types proposed by \citet[pp.~88--89]{fine_1991_PUF}. In his opinion, the four types of migration can be classified according to their relationship to time, ranging from short-term to long-term. The first type of migration concerns commuting between home and work, which is carried out periodically. The second type of migration characterizes seasonal movements related to economic activity, such as agricultural harvesting. In this case, individuals migrate temporarily so as to stay close to places where economic activity has temporarily increased during the season. When the latter ends, labor supply decreases and seasonal workers return home. The third category of migration is also temporary, but lasts longer. It describes temporary movements during a lifetime in connection with a change in activity. These first three types of migration are closely linked to the labor market.\footnote{The labor market is used to define a migrant. \citet[p.~374]{Shryock_1976_Demography} refer to the existence of increasing opportunity costs of distance from home to work, which can lead to a change of residence when it becomes too high.} The fourth type of migration is broader. It describes the permanent movements of individuals from the mountains to the countryside, and from the countryside to the cities. This study focuses on this fourth type of migration due to the data it uses.

This article contributes to the literature in historical demography by proposing a study of the French case in the 19th century. It presents observations on migration from generation to generation, drawing on rich individual data from the collaboration of hundreds of thousands of Internet users. The internal migration of the French is described through the prism of several spatial scales, ranging from the global level to a much finer level. The results show that the use of collaborative genealogy data not only makes it possible to recover known facts in the literature, but also to enrich them.

The rest of this article is organized as follows. \autoref{sec:data} presents the data. \autoref{sec:migration national} examines migration at the national level. \autoref{sec:short_long_dist_migration} takes a more detailed look at the types of migration by distance traveled. \autoref{sec:migration cities} proposes to examine movements between cities according to city size. \autoref{sec:migration micro} shows the possibilities offered by collaborative genealogy data to conduct studies at a very fine scale. Finally, \autoref{sec:migration} addresses the special case of migration to Paris, the capital of France.

\section{Data}\label{sec:data}
% ---------------------------------------- %

The analysis relies on collaborative data from a genealogy website, Geneanet.\footnote{\href{https://www.geneanet.org/}{https://www.geneanet.org/}} On that website, users looking for their ancestors collect information themselves, and fulfill their family tree.\footnote{The website's users can choose between publicly sharing their family tree or keep it private. We do not have access to the latter, so that our analysis only uses publicly shared family trees.} The information they provide about their ancestors corresponds to three types of events that can be found on civil and religious registers: birth, marriage, if any, and death. For each event, a date and a place can be fulfilled. Individuals in a tree are linked through their parents and spouses.

In this article, we focus on French migration in the nineteenth century. We follow the movements of people who were born in France\footnote{According to Geneanet, $40\%$ of their records relate to French data.} between 1800 and 1804,\footnote{As reminded by \citet{fleury_1958_Population}, this period corresponds to the uninterrupted recovery of death, marriage and death records in the whole of France.} and their offspring over three generations. Raw data includes $701~466~921$ observations. A substantial task of matching and data cleaning is, however, needed.\footnote{More details can be found in \autoref{appendix:construction}.} First, as each amateur genealogist constructs their family tree on the website, there are a lot of duplicated individuals in the raw data. We need to match individuals referring to a single person. Second, we need to clean the data to: ($i$) complete missing observations during the merger of trees, and ($ii$) correct some obvious mistakes. Completion of information can be done thanks to the large amount of data. Let us consider an example in which some information is missing, say the place of birth of an individual in a user's tree, but other information is present, \textit{e.g.}, the date of death. In the family tree of another user, the record that refers to the same individual provides details on the place of birth but lacks information regarding the date of death. Then, during the merger, we are able to complete the life events of the individual. The correction of mistakes also happens during the merger. If a user misspelled the first name of an individual, it is possible to correct it, provided that a majority of other users agree on another spelling. The complete methodology as well as the computer codes (written in R) are provided in an online annex available at the following address: \href{http://egallic.fr/Recherche/Genealogy/}{http://egallic.fr/Recherche/Genealogy/}.

\begin{figure}[htb]\centering
\caption{Distribution of Year of Birth in the Sample, by Generation.}\label{fig:distrib_naissances_gen}
\includegraphics[width = \textwidth]{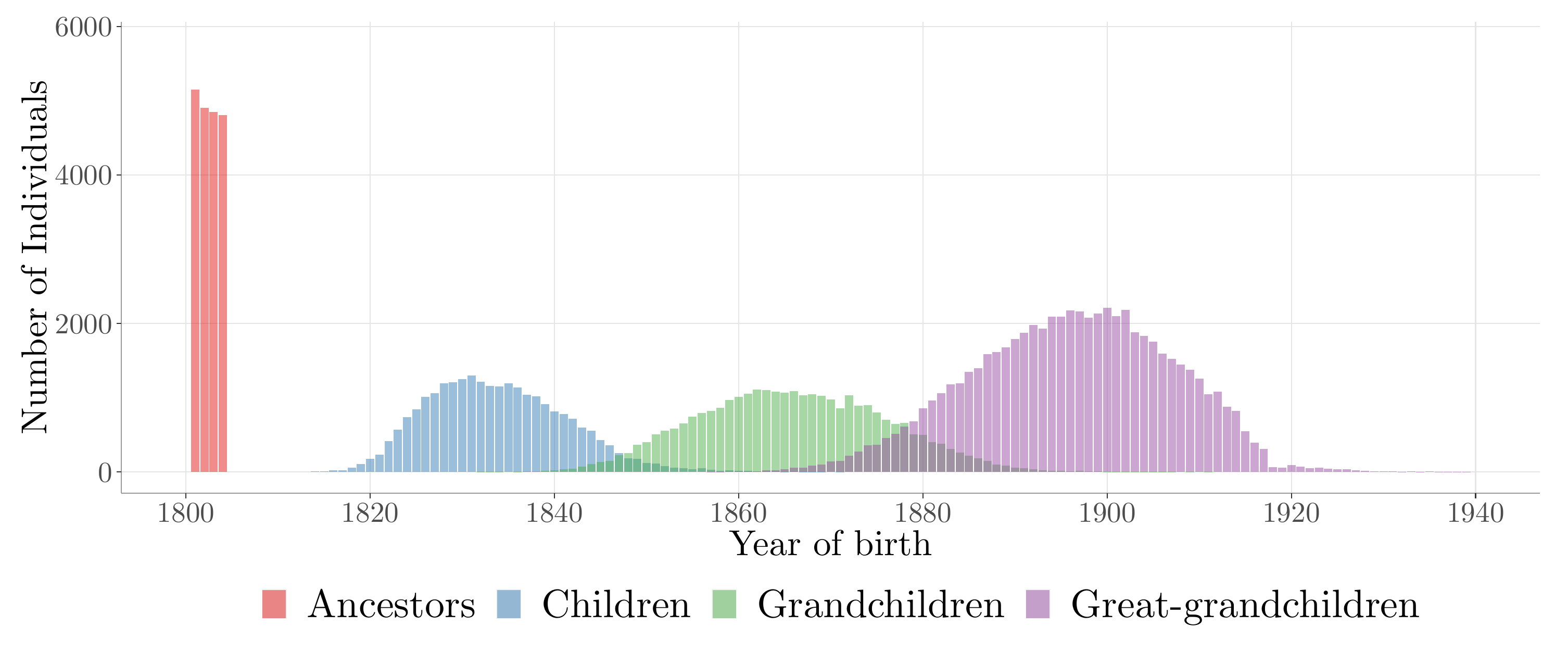}
%\begin{minipage}{\textwidth}
%		\vspace{1ex}
%		\scriptsize\underline{Note:} 
%\end{minipage}
\end{figure}

Once the family trees are matched and the data is cleaned, we proceed to a restriction on age. As we are interested in migration, we only consider people who are able to move. We thus keep people who survived at least until they are 16 years old and discard the rest of observations. We end-up with $25,485$ ancestors born in 1800 and 1804, $24,516$ children, $29,715$ grandchildren and $62,165$ rear grandchildren. The distribution of the year of birth, by generation, is shown in \autoref{fig:distrib_naissances_gen}.

% ---------------------------------------- %
\section{Migration at the National Level}\label{sec:migration national}
% ---------------------------------------- %

\begin{figure}[htb]\centering
\caption{Percentage of Descendants Born in a Department Different from that of their Ancestor, by Department.}\label{fig:naissance_meme_dep}
\includegraphics[width = \textwidth]{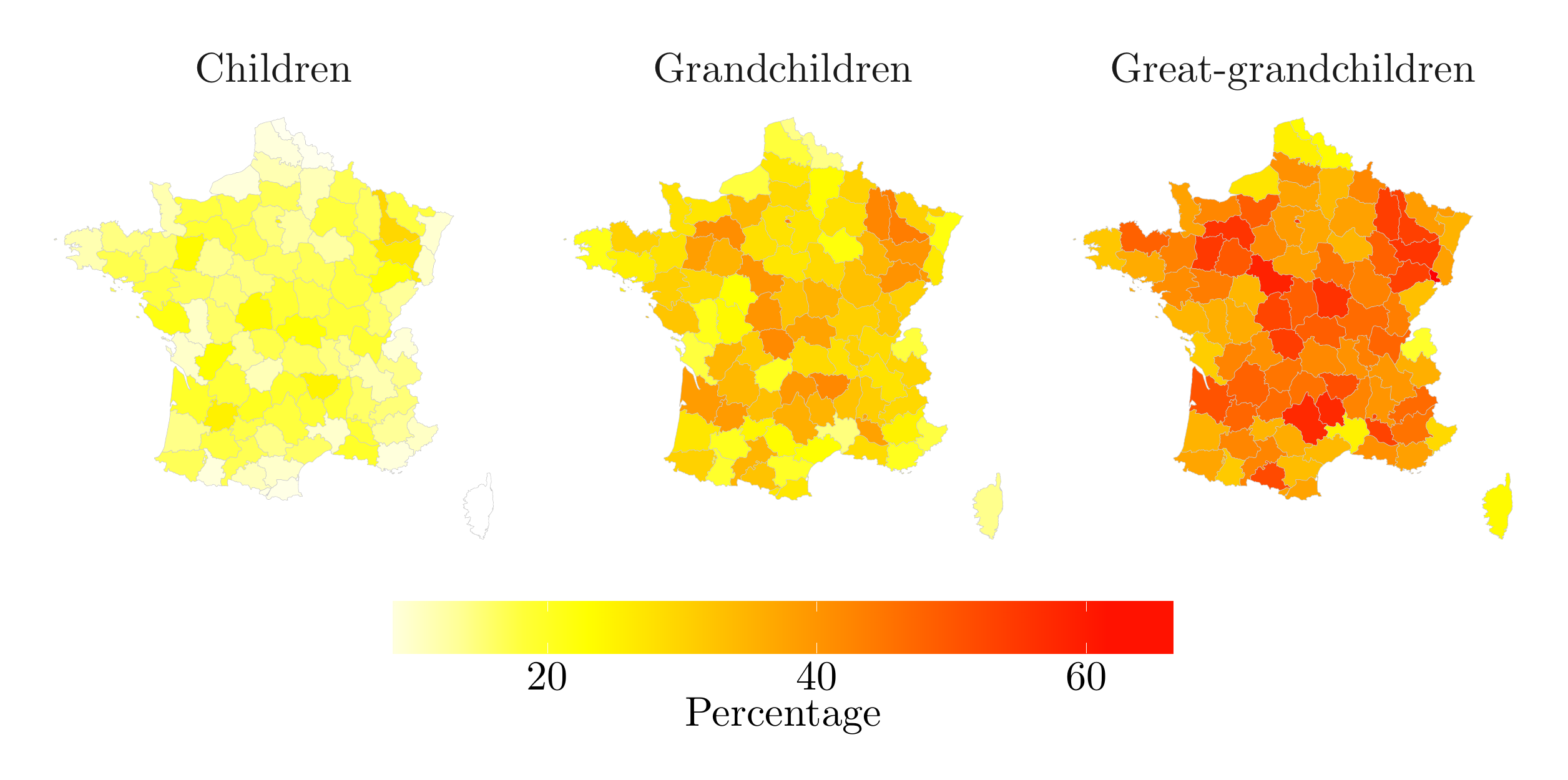}
\begin{minipage}{\textwidth}
		\vspace{1ex}
		\scriptsize\underline{Note:} These maps show the percentage of children (left), grandchildren (middle) and great-grandchildren (right) that were born in a department different from that of their ancestor who was born in France between 1800 and 1804.
\end{minipage}
\end{figure}

To get a global idea of regional heterogeneity of internal migrations, we rely on the information given regarding the place of birth of the individuals who were born in France between 1800 and 1804 and that of their descendants. More specifically, we extract the administrative area in which these individuals were born. These areas, which are called department, were established in 1790. They divide the European territory of France in 96 pieces.\footnote{Currently, there are 101 department in France, 96 of which are located in the European territory, so called `metropolitan France'.} Once we have obtained the birth department for both ancestors and their offspring, we compute the percentage of descendants for each ancestor, in each department, who were born in another administrative area. The resulting percentages are shown in \autoref{fig:naissance_meme_dep}, for each generation. The majority of the children of the individuals that make up the sample of ancestors were born in the same department as their ancestors. A smaller proportion is observed for the grandchildren and an even smaller one for the great-grandchildren. Significant regional differences can be seen from the third generation, \textit{i.e.}, the generation of the great-grandchildren. There are indeed proportions of children to be born in a department different from that of their ancestor relatively higher in the center of the country compared to the rest of France.

\begin{figure}[htb]\centering
\caption{Birthplace of Descendants Born Outside the Birth Department of their Ancestor.}\label{fig:distances_nais_heatmap}
	\begin{subfigure}{\textwidth}
		\includegraphics[width = \textwidth]{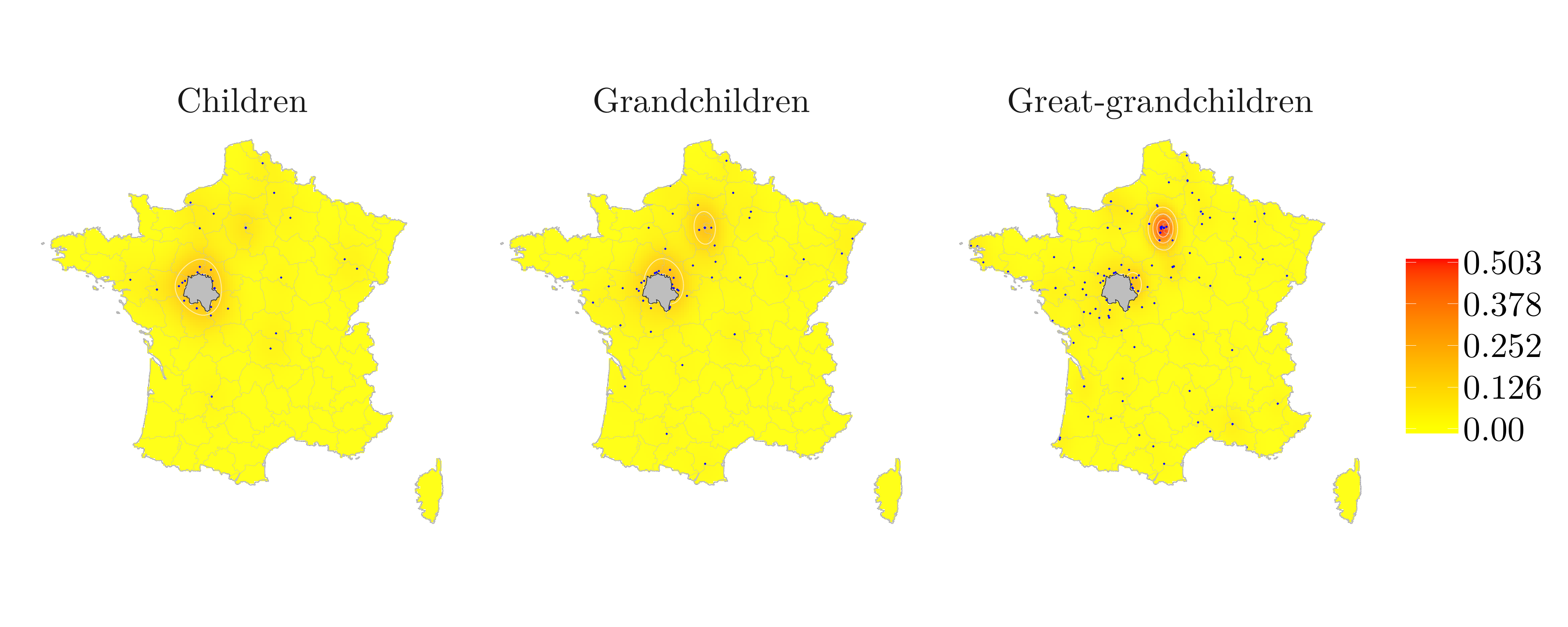}
		\vskip .5em
	    \caption{Indre-et-Loire}\label{subfig:distances_nais_heatmap_1}
	\end{subfigure}
	\begin{subfigure}{\textwidth}
		\includegraphics[width = \textwidth]{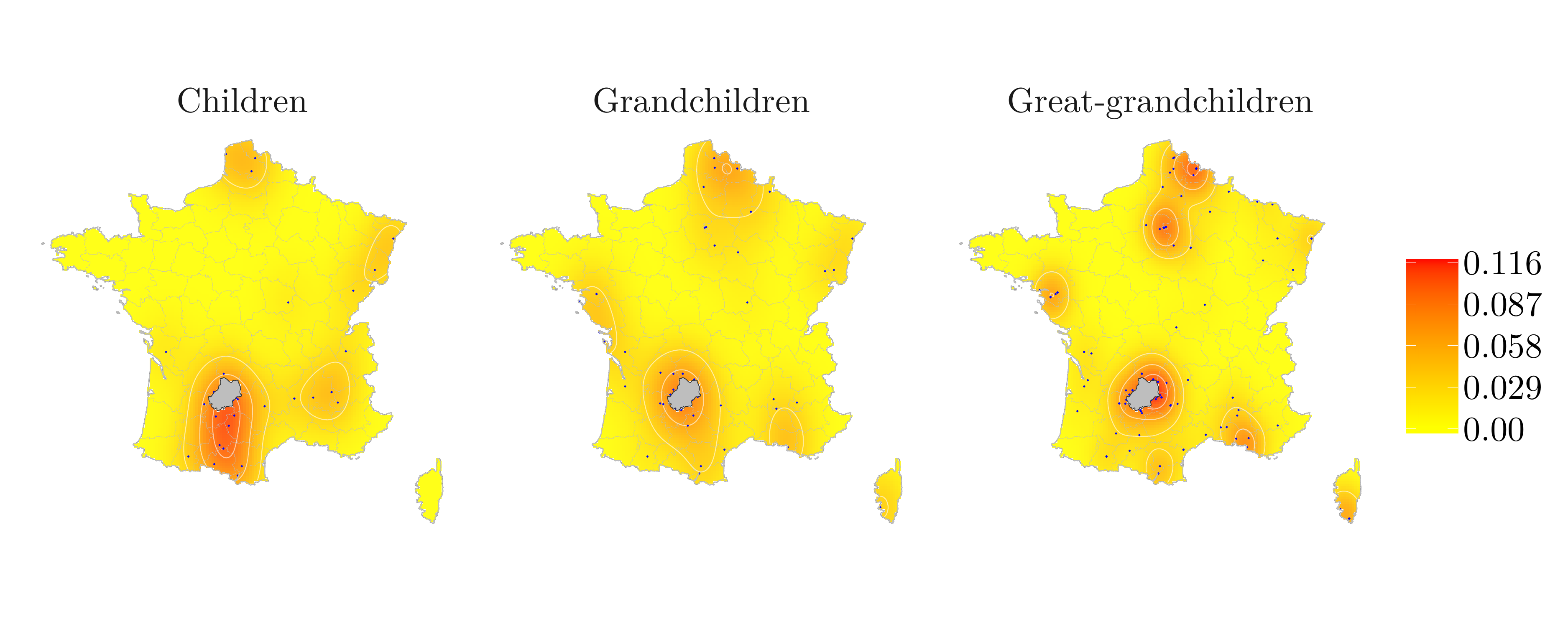}
		\vskip .5em
	    \caption{Lot}\label{subfig:distances_nais_heatmap_2}
	\end{subfigure}
\begin{minipage}{\textwidth}
		\vspace{1ex}
		\scriptsize\underline{Note:}
		These maps show the estimation of the density of the birthplace of the descendants of individuals born between 1800 and 1805 in two different French departments: Indre-et-Loire (top) and Lot (bottom). We excluded the descendants who were born in the same department as their ancestor.
\end{minipage}
\end{figure}

It is possible to focus only on the descendants who were born in a different department from that of their ancestor. For these movers, we extract the coordinate pairs (longitude, latitude) and estimate the density of births over the French territory, for each generation and each department. We use a `modified' Gaussian Kernel to estimate these densities.\footnote{As we are interested in internal migrations in France, we only focus on French territory. In addition, we are looking at the places of birth of the descendants who were born outside the department of their ancestor. So, when we compute the density of births for a specific generation and a specific department, we know that there will be no observations either in France or in the specific department. We therefore need to account for this in the density estimation to avoid some biases close to the frontiers. To correct these biases, we rely on the procedure explained in \citet{charpentier_2016_geoinformatica}.} We represent them using heat maps. An example is provided in \autoref{fig:distances_nais_heatmap}, for two French administrative divisions `Indre-et-Loire' (\autoref{subfig:distances_nais_heatmap_1}) and `Lot' (\autoref{subfig:distances_nais_heatmap_2}) which are located in the center-west and southwestern France, respectively. For each map, the corresponding birth department of the ancestors are filled with gray. As can be seen in both examples, most of the descendants were born not far from the department in which their ancestor was born. Some regions also stand out from the maps. This is the case for Paris, but not in the first generation. For the Lot, a hot spot can be seen in the North of France. It is possible that the presence of mining operations in the North of France caused the departure of certain individuals in this region. Still for the Lot, another economic pole stands out: the region of Marseille, for the second and third generation (grandchildren and great-grandchildren, respectively). Such a multitude of hot spots is not present for Indre-et-Loire, which reinforces the idea of regional differences as previously mentioned.

% ---------------------------------------- %
\section{Short and Long Distance Migrations}\label{sec:short_long_dist_migration}
% ---------------------------------------- %

\begin{figure}[htb]\centering
\caption{Migration Between Generations.}\label{fig:distance_moins_20_gen}
\includegraphics[width = \textwidth]{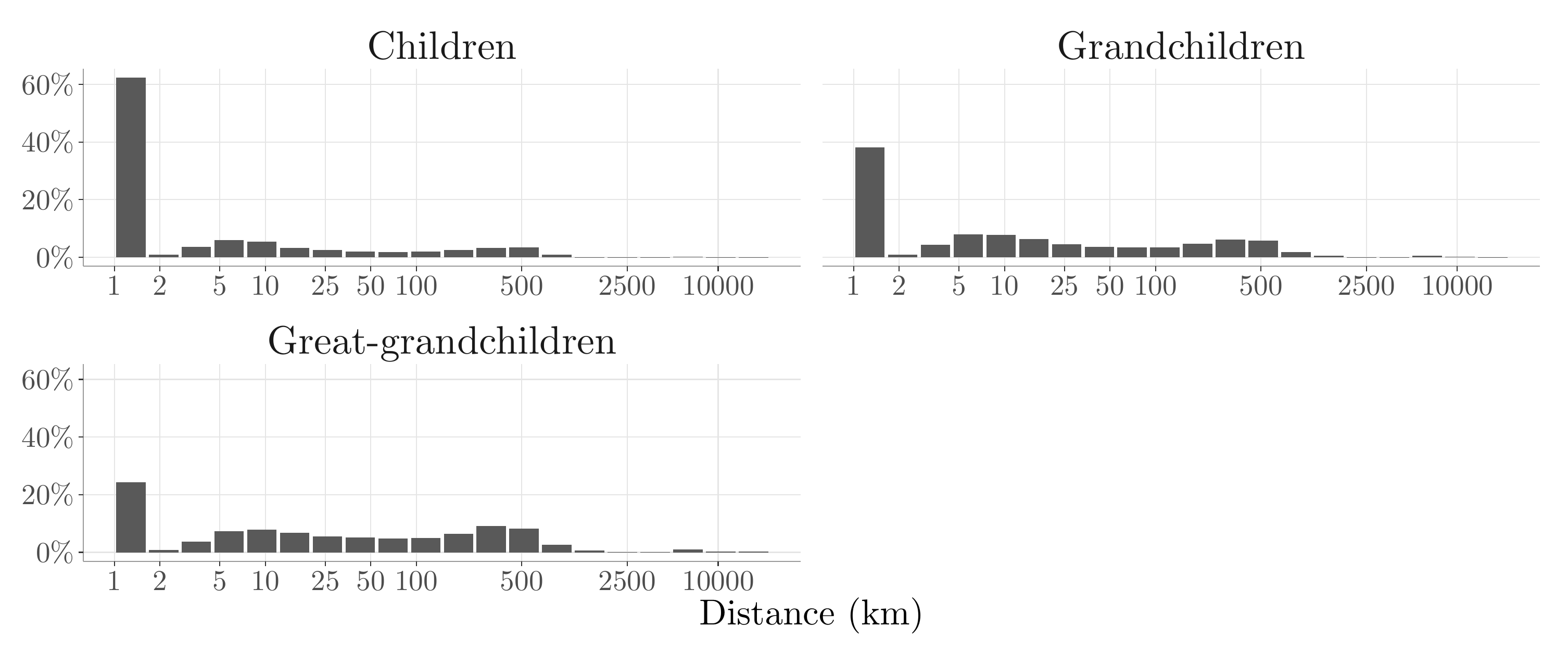}
\begin{minipage}{\textwidth}
		\vspace{1ex}
		\scriptsize\underline{Note:} The graphs show the distance distribution between the birthplaces of the individuals in the sample and the birthplaces of their closest parent, for each generation. Please note that the distance is plotted on a logarithmic scale.
\end{minipage}
\end{figure}

Rather than considering migration by the prism of regional changes, it is possible to leave the regional borders aside and to consider only the distances traveled from one generation to next. By using the geographic coordinates associated with the places of birth of the individuals in our database, it is possible to calculate the distances separating the places of birth of an individual from those of his or her descendants. We do it for the ancestors who were born in France between 1800 and their descendants. Contrary to what we did for the analysis of migration between departments, we consider births outside the territory of Metropolitan France. The distributions of distances between generations are provided in \autoref{fig:distance_moins_20_gen}. The graph shows that a large part of the descendants, regardless of their generation, were born at exactly the same place as their ancestors. Their share, however, decreases gradually over the generations, from $60\%$ for children to just over $20\%$ for great-grandchildren.

Like the results put forward by \citet{bourdieu_2000_Annales}, we observe bimodality in the distribution of distances among those who were not born in the same place as their ancestor. We can use this bimodality to disentangle between the short distance and the long-distance migrants. It is common in the literature to make such a distinction, although there is no consensus regarding the value that allows one to differentiate short-distance migrants from long-distance ones. While \citet{rosental_2004_ADH} uses a value of 25 km, \citet{bourdieu_2000_Annales} use 20 km and \citet{kesztenbaum_2008_HF} places the cursor at 17 km. All of them choose their value according to their data, so that it reflects the median distance traveled by migrants. In our sample, the median value for the individuals from the first generation, \textit{i.e.}, the children, is 20 km. We therefore use 20 km as the threshold value for separating short-distance and long-distance migrants. The idea behind separating migrants according to the distance they travel is based on economic and sociological arguments. As explained by \citet{kesztenbaum_2008_HF}, the higher the distance, the higher the costs, whether economic or not. \citet{Rosental_2006_FHS} argues that the social status of individuals is one of the factors that explain why people decide to move to another place. According to him, the most privileged, who were also those who received a better education, tend to travel long distances while people from modest background tend to travel short distance, and sedentariness represents an average.

In our sample, there is a strong sedentariness among the children, which amounts to $62\%$ (\autoref{tab:intergen_migration}). The share of sedentary in the ancestors decreases over the generations and falls to $38\%$ for the grandchildren and to $24\%$ for the great-grandchildren. At the same time, we note that long-distance migrations has increased significantly more than short-term migrations, from just $18\%$ for children to almost $50\%$ for great-grandchildren.

\begin{table}[ht]
\centering\scriptsize
\caption{Intergenerational Migrations in France, in Percentages.}\label{tab:intergen_migration}
\begin{tabular}{lrrr}
  \hline\hline\\[-.75em]
  & Children & Grandchildren & Great-grandchildren\\[.25em]
  \hline\\[-.75em]
	Sedentary & 62.17 & 38.06 & 24.17 \\ 
    Short-distance Migrants & 19.43 & 27.70 & 27.06 \\ 
  	Long-distance migrants & 18.40 & 34.24 & 48.77 \\ 
   \hline\hline
\end{tabular}
\begin{minipage}{\textwidth}
		\vspace{1ex}
		\scriptsize\underline{Note:} Sedentary were born in the same location as their ancestor; short-distance and long-distance migrants were born less than 20 km and more than 20 km from the birthplace of their ancestor, respectively.
\end{minipage}
\end{table}

%Looking at gender (\autoref{tab:gender_migration})
%
%
%
%
%\begin{table}[ht]
%	\centering
%	\caption{Type of migrations by gender for the individuals born in France between 1800 and 1804}\label{tab:prop_villes}
%	\begin{tabular}{lrr}
%	  \hline\hline
%	& Women & Men \\ 
%	  \hline
%	  Sedentary & 61.31 & 65.50 \\ 
%	  Short & 22.48 & 15.02 \\ 
%	  Long & 16.20 & 19.48 \\ 
%	   \hline\hline
%	\end{tabular}
%	\begin{minipage}{\textwidth}
%			\vspace{1ex}
%			\scriptsize\underline{Note:}
%			Each column in
%	\end{minipage}
%\end{table}

% ---------------------------------------- %
\section{Migration Between Small Cities and Larger Ones}\label{sec:migration cities}
% ---------------------------------------- %

A look at intergenerational migration can also be done at the city level. We consider that the place of birth of a newborn corresponds to the place where his or her parents are living. With recent data, this hypothesis would lead to important biases, as the cities in which the newborns are registered correspond to those of the maternities in which the newborns are born. However, for the period considered, \textit{i.e.}, the nineteenth century, a vast majority of birth deliveries take place at home. As reminded by \cite[p. 34]{fine_1991_PUF}, $98\%$ of deliveries still take place at home by the eve of the Second World War.

Information on population size of cities in terms of inhabitants can be added to classify the cities in different groups. Values for the year 1838 are provided at the scale of chief town and large cities \citep{insee_pop_1800}. We distinguish between four categories, as in \citet{fleury_1958_Population} and \citet{Blayo_1975_Population_mouv}:
\begin{itemize}
	\item large cities: more than $50,000$ inhabitants ;
	\item medium cities: between $10,000$ and $50,000$ inhabitants ;
	\item small cities: less than $10,000$ inhabitants ;
	\item extra-small cities: cities that are not included in the \textit{Statistique Générale de la France} data.
\end{itemize}

The number of large cities in 9: Paris, Lyon, Marseille, Bordeaux, Rouen, Toulouse, Nantes, Lille, and Strasbourg. The number of medium and small cities is 98 and 254, respectively. The remaining cities in our sample is classified as extra small. A vast majority of births occur in extra-small cities, as shown in \autoref{tab:prop_cities}. For the ancestors, \textit{i.e.}, those who were born in France between 1800 and 1804, $88\%$ of them were born in an extra small city. Their descendants were also mainly born in extra small cities. It can, however, be noted that the share slowly declines until $80\%$ by the great-grandchildren. In the meantime, the share of individuals that were born in a large or a medium city grows from $3.40\%$ for the ancestors to $7.63\%$ for the great-grandchildren. A similar growth is observed for medium cities. On the contrary, the share of individuals in each generation that were born in a small city remains almost unchanged. 

The decline observed in the proportion of grandchildren that were born in a small city can be linked to the phenomenon of rural exodus, which begins in the second half of the nineteenth century, according to \cite{lemercier_2000_Population}.

\begin{table}[ht]
\centering
\caption{Distribution of the Places of Birth of the Individuals in the Sample According to the Size of the City.}\label{tab:prop_cities}
\begin{tabular}{lrrrr}
  \hline\hline
Type of Municipality & Ancestors & Children & Grandchildren & Great-grandchildren \\ 
  \hline
	Large & 3.40 & 3.92 & 6.31 & 7.63 \\
  	Medium & 5.18 & 5.25 & 6.55 & 7.86 \\ 
	Small & 3.43 & 3.56 & 3.73 & 3.81 \\ 
	Extra Small & 87.99 & 87.27 & 83.41 & 80.70 \\ 
   \hline\hline
\end{tabular}
\begin{minipage}{\textwidth}
			\vspace{1ex}
			\scriptsize\underline{Note:}
			Each column gives, for a generation, the percentage of individuals in the sample that were born in each of the three categories of cities.
	\end{minipage}
\end{table}

If \autoref{tab:prop_cities} provides information on the place of birth of the descendants of the ancestors, it does not describe the origin and destination of migratory flows. To that end, we build a transition matrix for each generation that gives the percentage of descendants that were born in each city category, conditionally on the birth city of the ancestors. The transition matrix of each generation is provided in \autoref{tab:villes}. There is an overwhelming majority of sedentary among the descendants of people who were born in very small cities. However, the proportion of sedentary people declines slightly from one generation to the next, from $96\%$ for children to $85\%$ for great-grandchildren. These declines are offset by increases in proportions in medium and large cities. For ancestors born in large and medium-sized cities, the pattern is different. The share of sedentary among the descendants is relatively less important, compared to ancestors born in extra small cities. In the mean time, the percentage of descendants born in extra small cities grows over generations: it goes from $26\%$ for the children and rises to $50\%$ for the great-grandchildren. It should not be overlooked that the majority of individuals are born in extra small cities for each generation. The beginning of the rural exodus is not contradicted by these statistics, as a small share of $5\%$ of individuals that are born in an extra small city that migrate to a larger one represents more individuals than $50\%$ of persons born in a large city.

\begin{table}
	\centering\scriptsize
	\caption{Conditional Transitional Movements Between Large, Medium, and Small Cities, in Percentages.}\label{tab:villes}
	\begin{tabular}{llrrrrcrrrrcrrrr}
	\hline\hline\\[-.75em]
	&&\multicolumn{4}{c}{Children}&&\multicolumn{4}{c}{Grandchildren}&&\multicolumn{4}{c}{Great-grandchildren}\\
	 \cmidrule(lr){3-6} \cmidrule(lr){8-11} \cmidrule(lr){13-16}\\[-.75em]
	&&\multicolumn{1}{c}{L}&\multicolumn{1}{c}{M}&\multicolumn{1}{c}{S}&\multicolumn{1}{c}{XS}&&\multicolumn{1}{c}{L}&\multicolumn{1}{c}{M}&\multicolumn{1}{c}{S}&\multicolumn{1}{c}{XS}&&\multicolumn{1}{c}{L}&\multicolumn{1}{c}{M}&\multicolumn{1}{c}{S}&\multicolumn{1}{c}{XS}\\\\[.25em]
	\hline\\[-.75em]	&L&67.10&4.14&2.42&26.34&&52.67&7.47&2.93&36.92&&36.73&9.97&3.27&50.03\\
&M&5.75&67.97&1.96&24.32&&10.42&47.99&3.23&38.36&&14.02&33.85&3.02&49.11\\
&S&4.63&4.11&65.37&25.89&&10.21&7.64&37.46&44.69&&12.50&11.13&21.38&54.99\\
&XS&1.13&1.52&1.29&96.06&&3.62&3.92&2.39&90.07&&5.46&5.91&3.22&85.42\\
	\hline\hline
	\end{tabular}
	\begin{minipage}{\textwidth}
			\vspace{1ex}
			\scriptsize\underline{Note:} The tables contain a transitional matrix for each generation providing the frequency of births of descendants in a large (L) city, in a medium (M) city, in a small (S) city (S) or an extra small (XS) city (in column), conditional on the birthplace of the ancestors (in row).
	\end{minipage}
\end{table}

% ---------------------------------------- %
\section{Case Study: at the Micro Level}\label{sec:migration micro}
% ---------------------------------------- %

As the previous sections show, most individuals have remained sedentary from one generation to the next. Among people born in a different place from their ancestors, \autoref{sec:short_long_dist_migration} showed that for a significant proportion of them, the distance traveled remains small, within a radius of 20 km. This section pays particular attention to these individuals, focusing on migration at a micro level.

\begin{figure}[htb]\centering
\caption{Migration at a Finer Spatial Level.}\label{fig:micro_migration}
\includegraphics[width = \textwidth]{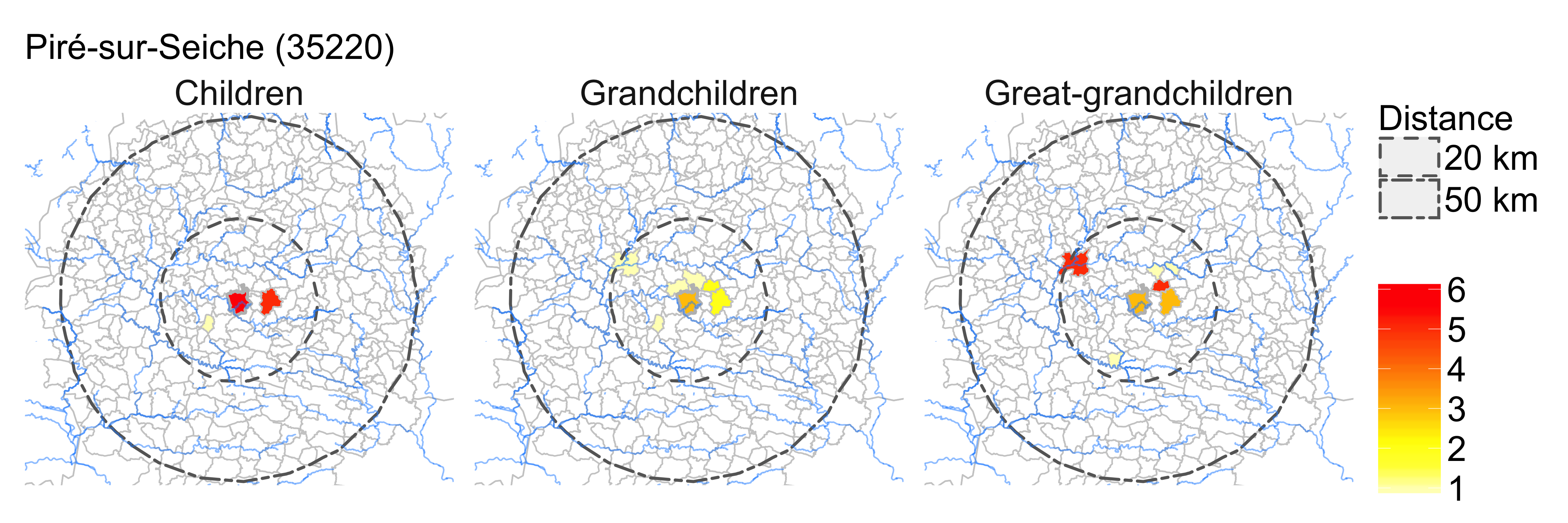}
\begin{minipage}{\textwidth}
		\vspace{1ex}
		\scriptsize\underline{Note:} The maps show, for each generation, in which city the descendants of individuals born in Pir\'e-sur-Seiche were born, limiting the analysis to a radius of 50 km.
\end{minipage}
\end{figure}

The idea developed here is to select a village, count the births in it for our ancestors, and look where the descendants were born. As the objective is to look at what is happening at a fine scale, only migrations over small distances are considered. We study the birthplaces of the descendants of a given village within a radius of 50 km. In most cases, since the majority of individuals are born in the same place as their parents, few movements are observed. In some other cases, however, notable movements between cities can be observed. This is the case of the city of `Piré-sur-Seiche', as illustrated in \autoref{fig:micro_migration}. Pir\'e-sur-Seiche is a small town in the department of `Ille-et-Vilaine', in the west part of France. On the maps, this commune is placed in the center and its boundaries are highlighted by a thicker gray line than for the other municipalities. Two dotted circles define a radius of 20 and 50 km, respectively. The map on the left shows the birthplaces of the children of individuals born between 1800 and 1804 in Pir\'e-sur-Seiche. While most of the children were born in the same commune, a significant proportion moved a few kilometers to a neighboring village called Bais. Among the grandchildren, the spread over nearby villages is also observed. It is interesting to note that a slightly more distant city (while remaining within a radius of 25 km) stands out for the generation of great-grandchildren. This town west of Pir\'e-sur-Seiche is called `Rennes'. It is a more important city both administratively and in terms of population. This shift from a small municipality to a large city can be seen in other examples, but not necessarily in all cases.

The analysis of migration at such a fine level is made possible by the individual nature of the data. It should be noted, however, that the initial restrictions in this article (intended to focus only on individuals born in the early 19th century and their descendants) only provide a glimpse of migration at the micro level. Indeed, since annual births in small villages are not very numerous, there are few movements observable on such a small scale. Collaborative genealogical data could, however, prove to be an excellent means of observing migrations on a very fine spatial scale, provided that a larger population of ancestors is considered than that chosen in this study.

% ---------------------------------------- %
\section{Migrations From and Towards the Capital}\label{sec:migration}
% ---------------------------------------- %

A special case in migration is that of Paris, the capital of France.

\begin{figure}[htb]\centering
\caption{Department of Birth of Ancestors of Descendants Born in Paris.}\label{fig:immigration_paris}
\includegraphics[width = \textwidth]{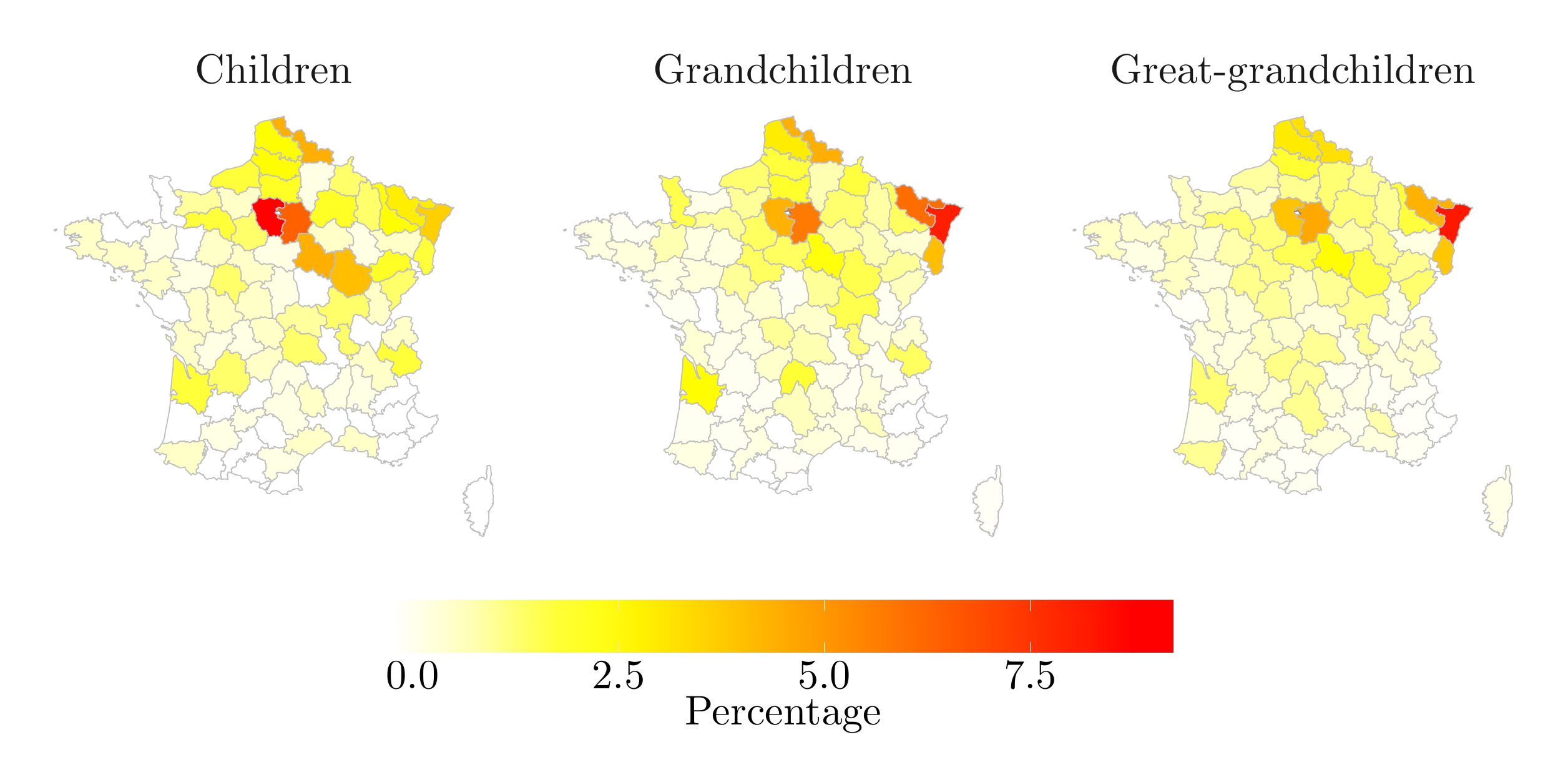}
\begin{minipage}{\textwidth}
		\vspace{1ex}
		\scriptsize\underline{Note:} The maps show, for each generation, the share of each department of birth of ancestors of descendants born in Paris, excluding ancestors born in Paris.
\end{minipage}
\end{figure}

First, we examine the origin of the ancestors of individuals born in Paris, at the departmental level. This is illustrated in \autoref{fig:immigration_paris}. For each generation (children on the left, grandchildren in the middle and great-grandchildren on the right), the maps show each department's share of the ancestors' origin. For clarity purposes, due to the strong sedentariness, the department of Paris has been removed from the calculation. To simplify the understanding, let us take an example. On the left map, the color of the northernmost department (the department called Nord) indicates a value of $4.69\%$. This reflects the fact that $4.69\%$ of the ancestors of the generation of children born in Paris, are from the North (excluding the ancestors born in Paris from the calculation). Interestingly, most of the ancestors of the individuals born in Paris were born relatively close to the capital. The further one moves away from the city of Paris, the less important the share of each department in the origin of the ancestors is.

\begin{figure}[htb]\centering
\caption{Percentage of Migrants to Paris.}\label{fig:immigration_paris_2}
\includegraphics[width = \textwidth]{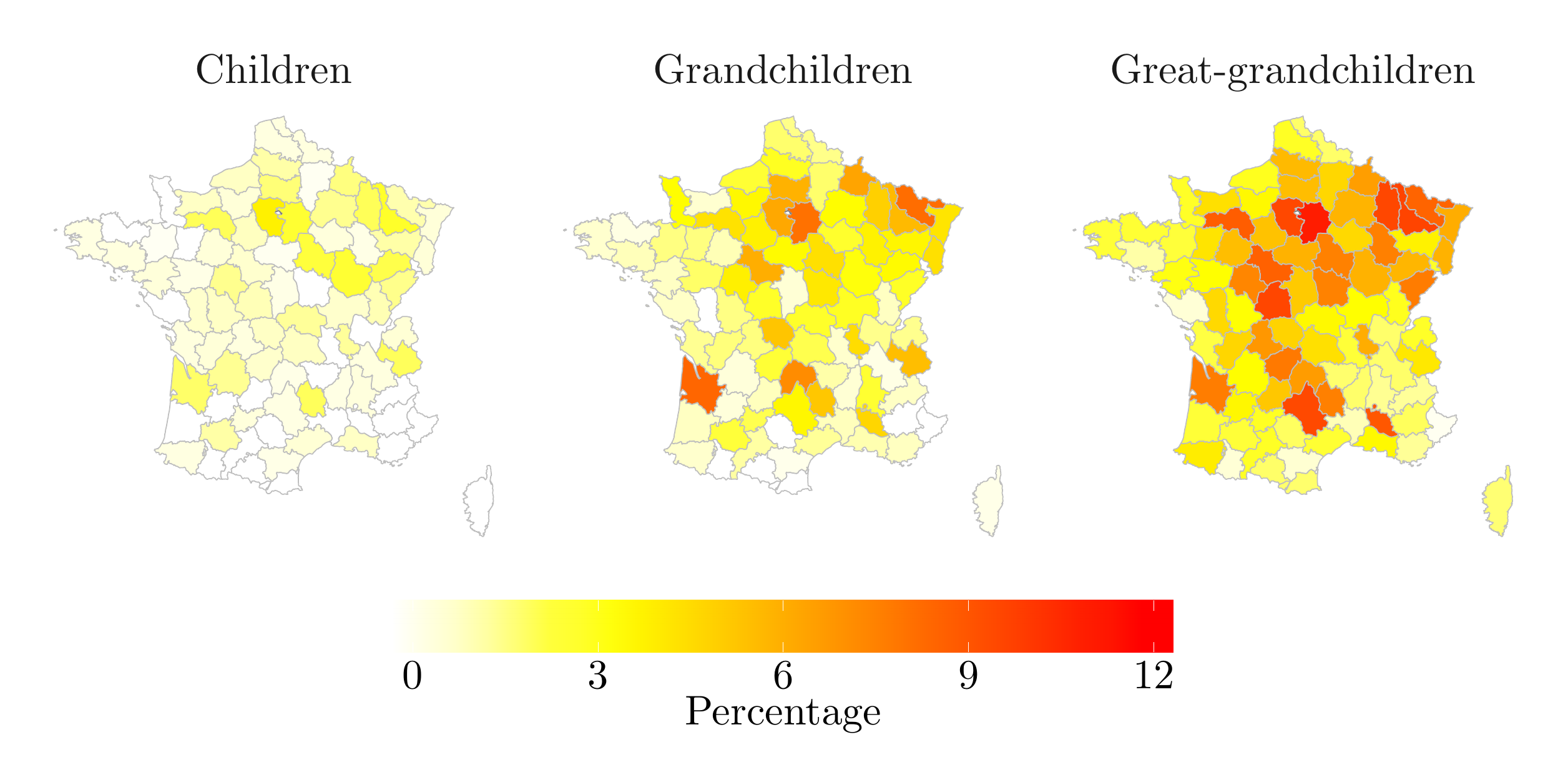}
\begin{minipage}{\textwidth}
		\vspace{1ex}
		\scriptsize\underline{Note:} The maps show the percentage of descendants who were born in Paris, for each generation and each department (with exception of Paris).
\end{minipage}
\end{figure}

Secondly, we look at the percentage of descendants born in Paris for each department. This is proposed in \autoref{fig:immigration_paris_2}.\footnote{Again, we exclude the department of Paris for visual concerns, so that the high percentage associated with this department does not inflate the scale.} We note that for individuals of the first generation, \textit{i.e.}, children, the percentage of descendants born in Paris is low for most departments. On average, just under $1\%$ of the children of individuals born between 1800 and 1804 were born in Paris per department. It was not until the next generation that we began to see more movement towards Paris. The departmental average of descendants in Paris goes from $0.88\%$ for children to $2.73\%$ for grandchildren. Some departments stand out for their higher values, notably `Gironde', where $8.65\%$ of small children were born in Paris. We also note that the southern and western departments of France are characterized by low values. Finally, for the generation of great-grandchildren, the departmental average of births in Paris rises to $4.69\%$, with a maximum of $11.2\%$ for `Seine-et-Marne'.

\begin{figure}[htb]\centering
\caption{Regional Proportions of Descendants Born in Paris According to the Distance Separating Each department to the Capital.}\label{fig:p_prop_paris_distance}
\includegraphics[width = \textwidth]{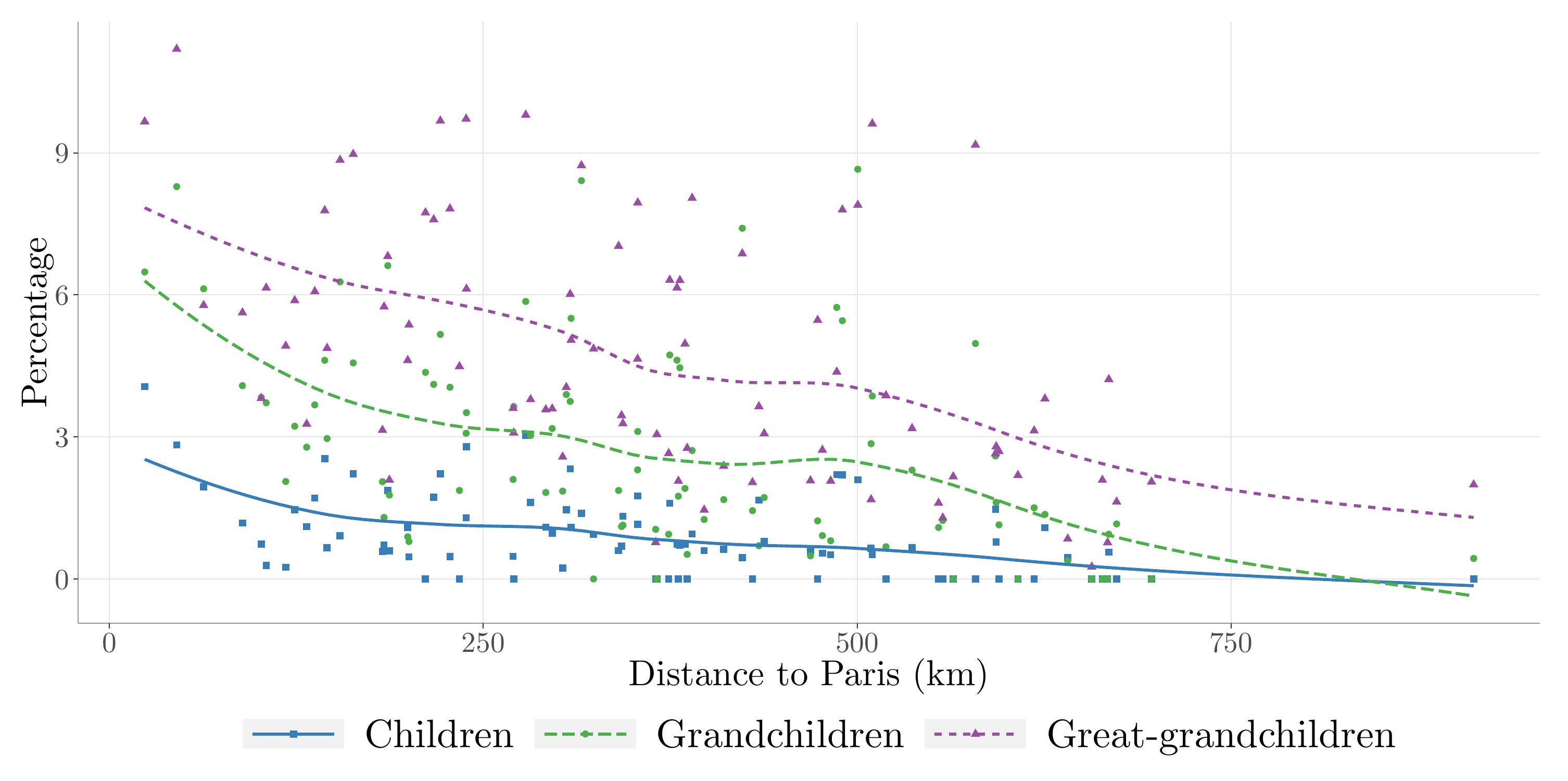}
\begin{minipage}{\textwidth}
		\vspace{1ex}
		\scriptsize\underline{Note:} Each dot represents the share of descendants of a department (green for children, turquoise for grandchildren, mauve for great-grandchildren) born in Paris. The distance of a department to Paris chosen here is that which separates the center of this department from that of the Paris region. The lines correspond to a \textit{loess} smoothing.
\end{minipage}
\end{figure}

Again, these maps seem to have a distance effect: the closer the department is to Paris, the higher the proportion of descendants to be born in the capital.  This is in fact what we can observe in \autoref{fig:p_prop_paris_distance}, for which we plot the percentage of descendants in Paris as a function of the distance from the original departments of the ancestors to the capital.\footnote{The distance between the regions and Paris is calculated using the distance separating the centroids from the regions.} Negative trends support the idea that the closer the departments are to Paris, the higher the proportion of descendants to be born there, whatever the generation. Moreover, the more we advance in the generations, the larger the average proportions are.

%\subsection{The distance effect}

To continue with the idea of the effect of distance on migration in Paris, it is possible to look at what is happening on a smaller spatial scale, now looking at migration out of Paris. We specifically examine transient movements between Paris, its crown and the rest of France. The crown is defined here as the area around Paris within a radius of 20 km from its limits. We report in \autoref{tab:villes_paris} conditional transition matrices for each generation, which indicate, conditionally to the place of birth of the individuals (in Paris or in its Crown), the place of birth of the descendants (in Paris, in its Crown or elsewhere). When the ancestors were born in Paris, the proportion of their descendants also born in Paris is significant, but decreases from generation to generation, going from $79.79\%$ of children to $43.45\%$ of great-grandchildren. While an increasing proportion of the descendants were born in the Paris suburbs, an equally growing, but relatively larger, proportion was born beyond the 20 km surrounding Paris. While one observes progressive movements of departures from the capital, opposite flows are observed in the crown. While only $4.36\%$ of the children of individuals born in the Paris suburbs were born in the capital, this percentage rose in subsequent generations to $11.79\%$ for great-grandchildren.

\begin{table}[htb]
	\centering\scriptsize
	\caption{Conditional Transitional Movements Between Paris and its Suburbs.}\label{tab:villes_paris}
	\begin{tabular}{llrrrcrrrcrrr}
		\hline\hline\\[-.75em]
		&&\multicolumn{3}{c}{Children}&&\multicolumn{3}{c}{Grandchildren}&&\multicolumn{3}{c}{Great-grandchildren}\\
		\cmidrule(lr){3-5} \cmidrule(lr){7-9} \cmidrule(lr){11-13}\\[-.75em]
	 &&\multicolumn{1}{c}{Paris}&\multicolumn{1}{c}{Suburbs}&\multicolumn{1}{c}{Other}&&\multicolumn{1}{c}{Paris}&\multicolumn{1}{c}{Suburbs}&\multicolumn{1}{c}{Other}&&\multicolumn{1}{c}{Paris}&\multicolumn{1}{c}{Suburbs}&\multicolumn{1}{c}{Other}\\
	\hline\\[-.75em]
Ancestors&Paris&74.79&3.56&21.64&&55.04&9.07&35.89&&43.45&8.98&47.58\\
&Suburbs&4.36&79.65&15.99&&6.77&68.42&24.81&&11.79&58.19&30.02\\
	\hline\hline
	\end{tabular}
	\begin{minipage}{\textwidth}
		\vspace{1ex}
		\scriptsize\underline{Note:}
		The table contains a transition matrix for each generation that gives the frequency of births in Paris, in its suburbs (<20 km) or further away (in column), conditional on the birthplace of the ancestors (in line).
	\end{minipage}
\end{table}

\clearpage

%\section{\refname}
%\footnotesize\singlespacing\linespread{1.5}
%\bibliographystyle{apalike}
\bibliographystyle{apalike}
\bibliography{bibCG}
%\normalsize\doublespacing

%------------------------------------------------------------------------%
%
%								APPENDIX
%
%------------------------------------------------------------------------%

\newpage

\appendix

% ---------------------------------------- %
\section{Construction of Family-Tree From the Data}\label{appendix:construction}
% ---------------------------------------- %

\subsection{From Ascending to Descending Genealogy}

Genealogist mainly uses an ascending strategy to research information on their ancestors \citep{Brunet_2015}. They start with an individual and adopt a bottom to the top technique as they look for the ancestors of that individual. The data we get from Geneanet correspond to this ascending system: the record of each individual contains a set of identifiers that allows us to connect him or her to their parents, whenever possible. However, as we want to study the geographical dispersion of the offspring of people who were born between 1800 and 1804, it is wiser to adopt a descending strategy which consists in identifying the descendants of our ancestors. An example is provided in \autoref{subfig:tree_base}. On the left, we represented two observations from the free of an amateur genealogist. These two observations refer to  \{Charles M\'elanie Abel\}-\{Hugo\} and \{Ad\`ele\}-\{Hugo\} who were born in Paris in 1826, and 1830, respectively. Both records point to the same parents: \{Victor Marie\}-\{Hugo\} born in Besançon in 1802 and \{Ad\`ele Julie\}-\{Foucher\} born in Paris in 1803. Since we want to examine the migration from a generation to the next, we need to retrieve the children of the individuals, not their parents. In our example with the Hugo family, we want to know that Victor Hugo had some children called Charles and Ad\`ele,\footnote{Actually, Victor Hugo had more children, but they are not represented in \autoref{subfig:tree_base} for simplicity.} as illustrated in \autoref{subfig:tree_wanted}.

\def \largeurnodes {1.75cm}

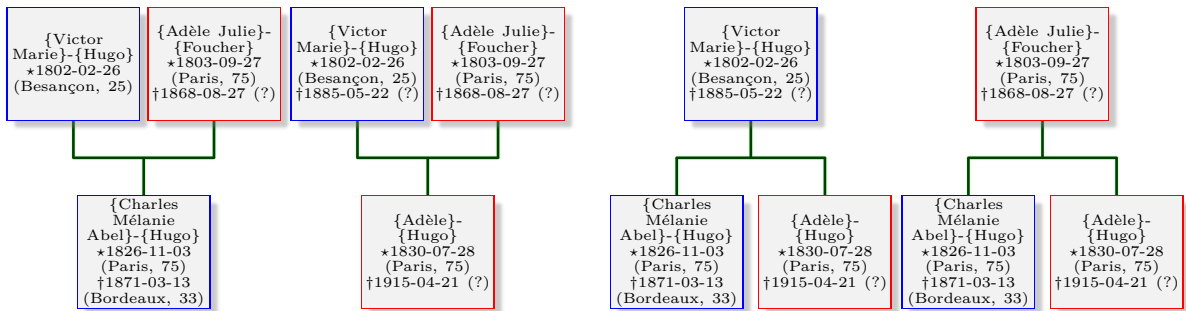
\begin{figure}[ht]
	\centering
	\caption{Examples of an Extract of a User's Family-Tree.}\label{fig:tree}
	\begin{subfigure}{.49\textwidth}
	    \begin{tikzpicture}
			\genealogytree[template=signpost,
			box={size=tight},
			node size=\largeurnodes,]
			{
				child{
					    g[id=pE, male]{\scriptsize \{Victor Marie\}-\{Hugo\}\\$\star$1802-02-26 (Besançon, 25)}
					    p[id=pF, female]{\scriptsize \{Adèle Julie\}-\{Foucher\}\\$\star$1803-09-27 (Paris, 75)\\$\dagger$1868-08-27 (?)}
						child{
							g[id=pE, male]{\scriptsize \{Charles Mélanie Abel\}-\{Hugo\}\\$\star$1826-11-03 (Paris, 75)\\$\dagger$1871-03-13 (Bordeaux, 33)}
						}
				}
			}
			\end{tikzpicture}
			\begin{tikzpicture}
			\genealogytree[template=signpost,
			box={size=tight},
			node size=\largeurnodes,]
			{
				child{
					    g[id=pE, male]{\scriptsize \{Victor Marie\}-\{Hugo\}\\$\star$1802-02-26 (Besançon, 25)\\$\dagger$1885-05-22 (?)}
					    p[id=pF, female]{\scriptsize \{Adèle Julie\}-\{Foucher\}\\$\star$1803-09-27 (Paris, 75)\\$\dagger$1868-08-27 (?)}
						child{
							g[id=pE, female]{\scriptsize \{Adèle\}-\{Hugo\}\\$\star$1830-07-28 (Paris, 75)\\$\dagger$1915-04-21 (?)}
						}
				}
			}
			\end{tikzpicture}
		\vskip .5em
	    \caption{Raw Data: Two Individuals Who Share the Same Parents.}
	    \label{subfig:tree_base}
	\end{subfigure}
	\begin{subfigure}{.49\textwidth}
			\begin{tikzpicture}
			\genealogytree[template=signpost,
			box={size=tight},
			node size=\largeurnodes,]
			{
				child{
					    g[id=pE, male]{\scriptsize \{Victor Marie\}-\{Hugo\}\\$\star$1802-02-26 (Besançon, 25)\\$\dagger$1885-05-22 (?)}
						child{
							g[id=pE, male]{\scriptsize \{Charles Mélanie Abel\}-\{Hugo\}\\$\star$1826-11-03 (Paris, 75)\\$\dagger$1871-03-13 (Bordeaux, 33)}
						}
						child{
							g[id=pE, female]{\scriptsize \{Adèle\}-\{Hugo\}\\$\star$1830-07-28 (Paris, 75)\\$\dagger$1915-04-21 (?)}
						}
				}
			}
			\end{tikzpicture}
			\begin{tikzpicture}
			\genealogytree[template=signpost,
			box={size=tight},
			node size=\largeurnodes,]
			{
				child{
					    g[id=pF, female]{\scriptsize \{Adèle Julie\}-\{Foucher\}\\$\star$1803-09-27 (Paris, 75)\\$\dagger$1868-08-27 (?)}
						child{
							g[id=pE, male]{\scriptsize \{Charles Mélanie Abel\}-\{Hugo\}\\$\star$1826-11-03 (Paris, 75)\\$\dagger$1871-03-13 (Bordeaux, 33)}
						}
						child{
							g[id=pE, female]{\scriptsize \{Adèle\}-\{Hugo\}\\$\star$1830-07-28 (Paris, 75)\\$\dagger$1915-04-21 (?)}
						}
				}
			}
			\end{tikzpicture}
			\vskip .5em
			\caption{What We Want : the Descendants of Each Parent.}\label{subfig:tree_wanted}
	\end{subfigure}
\end{figure}

\subsection{Data Cleaning}

As each genealogist from the website creates its own family tree, there are a lot of duplicated individuals in the raw data. An ancestor can appear in different user's trees with more or less information completed regarding his or her life events. A substantial matching work needs to be done. During this matching stage, we both clean, complete and correct the data. In a nutshell, as the volume of observations is quite large ($700$ million rows) and as our computer processing capabilities are limited, we split the sample into subsamples, one for each French department.\footnote{Since 1990, French territory is divided in administrative regions called `\textit{d\'epartements}'. Currently, there are 101 departments, 96 of which are located in the European territory. We focus on these 96 European French departments.} For each department, we identify the duplicated individuals using an algorithm and we reunite them. Once we have applied our algorithm to each department, we regroup the data in a single sample. We screen the data to identify the remaining duplicated individuals and merge them.

Our algorithms that aim at identifying and merge duplicated individuals work in a six-step procedure.
\begin{enumerate}
	\item We apply a very simple procedure which consists in grouping individuals that share the same following characteristics: last name, first names, gender, date of birth, last name of the mother, last name of the father.
	\item We deal with small spelling mistakes in last and first names. We consider that two people who were born in the same department, the same year, who have similar names (e.g., $\text{\{Ma\textbf{tt}hieu Paul\}-\{Henri\}}$ or $\text{\{Ma\textbf{t}hieu Paul\}-\{Henri\}}$), and whose parents' names are also close, can be identified as the same person. We rely on a string distance measure to evaluate how close two names are to each other. The measure we use is the cosine distance (see \citet{cohen2003comparison} for more details).
	\item We correct gender input errors, using the most frequent declared gender when there is a mismatch between information from different user's family trees.
	\item We correct the dates, using the same method as in the third step. Some dates are incomplete in some family trees, where only the year and not the month and day are provided. We complete these dates whenever it is possible, by looking at the information contained in other genealogist's trees.
	\item We proceed as in the first step, but this time, instead of looking at the complete set of first names, we only look at the first one that is provided.
	\item We list the brothers and sisters of any individual from each user's family tree. Among these sororities, we check whether some persons share the same first name and are born or dead the same day. If we find some, we consider that they can be merged as a single person.
\end{enumerate}

\end{document}